\def\APJ{\em Ap. J.}
\def\AP{\em Astropart.  Phys.}
\def\AA{\em Astr. and Astroph.}
\def\APJL{\em Ap. J. Letters}
\def\ARAA{\em Ann. Rev. Astr. Ap.}
\def\NAT{\em Nature}

\def\NIM{\em Nucl. Instrum. Methods}
\def\NIMA{{\em Nucl. Instrum. Methods} A}
\def\NIMBPROC{{\em Nucl. Instrum. Methods (Proc. Suppl.)} B}
\def\NPB{{\em Nucl. Phys.} B}

\def\PRL{\em Phys. Rev. Lett.}
\def\PL{\em Phys.  Lett.}
\def\PRD{{\em Phys. Rev.} D}

\def\be{\begin{equation}}
\def\ee{\end{equation}}
\def\bea{\begin{eqnarray}}
\def\eea{\end{eqnarray}}

\documentstyle[frascatiphys,epsfig]{article}

\begin{document}

\title{ SEARCH FOR ANTIMATTER IN SPACE WITH THE ALPHA MAGNETIC SPECTROMETER}

\author{R. Battiston\\ \em{Dipartimento di Fisica and Sezione INFN}\\ \em{Via Pascoli, 
Perugia, 06100 , Italy} \\ E-mail:battisto@krenet.it\\  AMS   Collaboration}

\maketitle


\abstract{ The Alpha Magnetic Spectrometer (AMS) is a state of the art 
particle physics experiment  for the  extraterrestrial  study of antimatter, matter and missing matter. 
AMS successfully  completed the precursor  STS91 Discovery  flight (June 2nd-12th, 1998), 
completing 152 orbits at  $\pm 52^o$ of latitude   and about  400 km of height, 
 collecting more than  100 million CR  events. In this paper we report on the first flight  experience
and we present preliminary results   on the search for  nuclear antimatter. No antimatter 
nuclei with $Z\geq2$ were detected. We obtain  a  model dependent upper limit  on 
  $\Phi_{\bar{He}}/\Phi_{He}$ of $<1.14\ 10^{-6}$.
In the rigidity  region between $1.6$ to $20$ GV we obtain a model independent, conservative 
 upper limit on  $\Phi_{\bar{He}}/\Phi_{He}$  of $<1.7\ 10^{-6}$ and of 
 $<2.8\ 10^{-5}$ for $Z>2$, improving  the results of previous searches performed 
 with stratospheric   balloons}

\endabstract
\vskip 4cm
Invited talk at XIIIth Rencontres de Physique: Results and Perspectives in Particle Physics, La
Thuile, Aosta Valley, February 22-27, 1999.

\clearpage 

\section{Introduction}
   The disappearance of cosmological antimatter and the pervasive presence of dark matter 
are two of  the greatest puzzles in the current  understanding of  our universe.

\begin{figure}[htb]
\begin{center}
\mbox{\epsfig{file=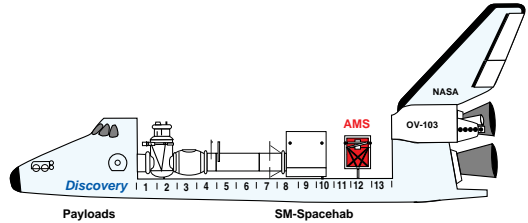,width=7 cm}}
\caption{AMS on STS 91 (Discovery), June 2nd - 12th, 1998.  \label{fig:STS91}}
 \end{center}
\end{figure}

The Big Bang model  assumes that, at its  very beginning,  half  of the universe
 was made out  of antimatter. The validity of this model is based
on three key experimental observations: the recession of galaxies (Hubble expansion), 
the highly isotropic cosmic microwave background and  the relative abundances of
 light isotopes.  However, a fourth basic observation, the presence of cosmological 
 antimatter somewhere in the universe, is missing.  Indeed measurements of the 
intensity  of gamma ray flux in the MeV region exclude the presence of a significant
 amount of antimatter up to the scale of the local  supercluster 
of galaxies  (tens of Megaparsecs). Antimatter should have been 
destroyed immediately  after the Big Bang  due to a mechanism  creating a matter-antimatter asimmetry
 through  a large violation  of CP and the baryon 
number~\cite{sk}. Alternatively   matter and 
antimatter  were separated  into different region
 of space, at scales larger than superclusters~\cite{br}. Other  possibilities
have also been recently suggested~\cite{kl}. 
 All efforts to reconcile the  
the absence of antimatter  with cosmological models 
 that do not require new physics failed~\cite{st}.

We are  currently unable to explain the fate of half of the  
baryonic matter present at the beginning of our  universe.

\begin{figure}[htb]
\begin{center}
\mbox{\epsfig{file=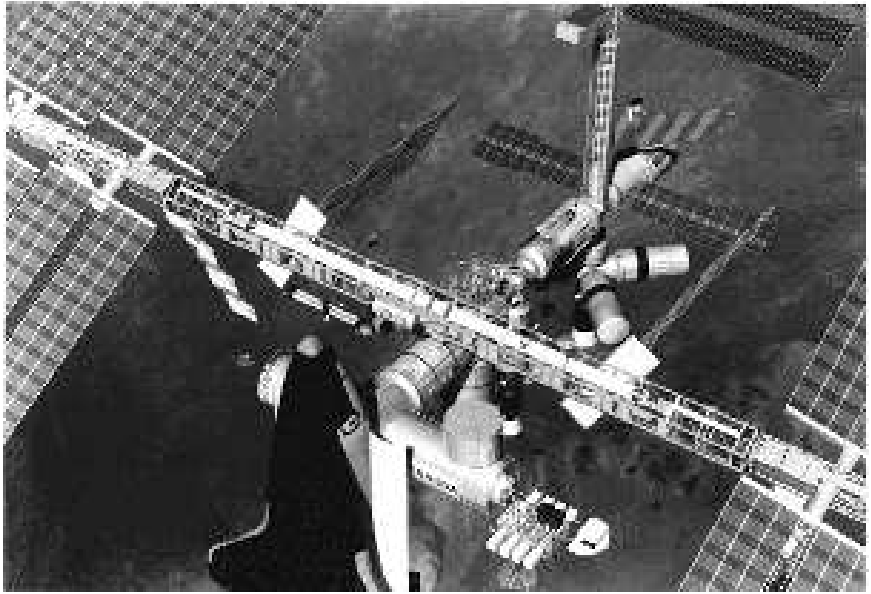,width=11.5cm}}
\caption{The  International Space Station Alpha; AMS will be installed 
on the left side of the main truss.  \label{fig:station}}
\end{center}
\end{figure}

Rotational velocities in spiral galaxies and dynamical effects in galactic clusters
provide us convincing evidence that, either Newton laws break down at scales of galaxies  
 or, more likely,  most  of our universe consists of non-luminous 
 (dark) matter~\cite{zw}. 
There are several dark matter candidates (for a recent review see~\cite{sa}).
They are commonly classified  as ''hot'' and ''cold'' dark matter, 
depending on their relativistic  properties at the time of decoupling from normal
 matter in   the early universe. As an example, light neutrinos are  
 obvious candidates for ''hot'' dark matter while  Weakly Interacting
 Massive Particles (WIMP's) like the lightest SUSY  particle (LSP)  are 
often considered as    plausible ''cold'' dark matter candidate~\cite{el}. 
Even  the recent results suggesting   a  positive  cosmological constant~\cite{ch} reducing the 
amount of  matter in the universe, confirm  
the dominance of dark matter over baryonic matter. 

	We are then    unable to explain the origin of most
 of the mass of our universe.

To address these two fundamental questions in astroparticle physics
  a  state of the art particle  detector, the Alpha Magnetic Spectrometer (AMS)~\cite{ahl}
has been approved  in 1995  by NASA to operate on  the
 International Space Station (ISS).

	AMS has successfully  flown on the  precursor flight (STS91, Discovery, June 2nd  1998,
 Figure~\ref{fig:STS91}), 
and it is approved  for a three year long exposure on the International Space Station (ISS),
 (Figure~\ref{fig:station}),
 after its installation  during Utilization Flight n.4, now  scheduled in 2004. 
AMS  has been proposed   and has been   built by  an international collaboration 
coordinated   by    DoE,  involving  China, Finland, France, Germany, Italy, Portugal, Rumenia,  
Russia, Spain, Switzerland,  Taiwan  and US. 

	In this conference we report on  the operation of AMS during the precursor flight and 
we give   preliminary results on the search for nuclear antimatter.



\section{AMS design principles and operation during the Shuttle flight}
\par
  Search of antiparticles  requires the capability to
identify  with
 the highest degree of confidence, the  type of particle traversing 
the experiment by measuring its mass,   the absolute value  and the sign of its electric charge.
This can be achieved through repeated 
measurements of the particle momentum (Magnetic Spectrometer), velocity 
(Time of Flight, Cerenkov detectors)
and energy deposition (Ionization detectors).

\begin{figure}[htb]
\begin{center}
\mbox{\epsfig{file=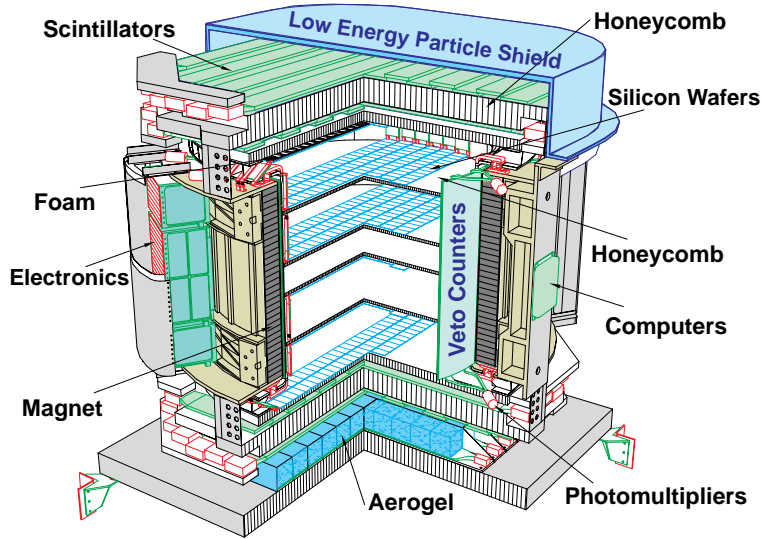,width=10 cm}}
\caption{AMS configuration for the June 1998  Shuttle  precursor  flight (STS91);
 about $45\%$ of the Silicon Tracker  was equipped during this mission.
\label{fig:AMSprecursor}}
\end{center}
\end{figure}

AMS 
configuration on the precursor flight 
is shown in Figure~\ref{fig:AMSprecursor}.
It consists on a large acceptance magnetic spectrometer ($0.6\ m^2 sr$) based on  
 a  permanent Nd-Fe-B Magnet , surrounding   a six layer high
 precision Silicon Tracker     and sandwiched between the four planes 
of the  Time of Flight scintillator system ($ToF$).



 A scintillator Anticounter system,
 located on the  magnet inner  wall
 and a  aerogel threshold Cherenkov   detector  ($n=1.035$), 
complete the experiment. A thin shield on the top and the bottom sides absorb low energy particles
present on  the Earth radiation belts. The detector works in  vacuum:
 the amount of material in front of the $ToF$ is about $1.5\ g/cm^2$ and $4\ g/cm^2$ 
in front of the Tracker.

The magnet is based on recent advancements in permanent magnetic material  technology 
 which made it possible to use very high grade Nd-Fe-B to construct 
a permanent magnet with $BL^2=\ 0.15\  Tm^2$ weighting $\le 2$ tons.
The magnet has a cylindrical shape with $80\ cm$ of  height  and an internal  diameter
 of  $100\ cm$.
A charged particle traversing the  spectrometer  experiences
 a dipole field orthogonal to the cylinder axis: it triggers the experiment  through the 
$ToF$  system (planes $S1$ to $S4$) which also    measures the  particle  velocity  ($\beta$) 
with a typical resolution of
$  105\ ps$ over a distance of $\sim \ 1.4\ m$.

\begin{figure}[htb]
\begin{center}
\mbox{\epsfig{file=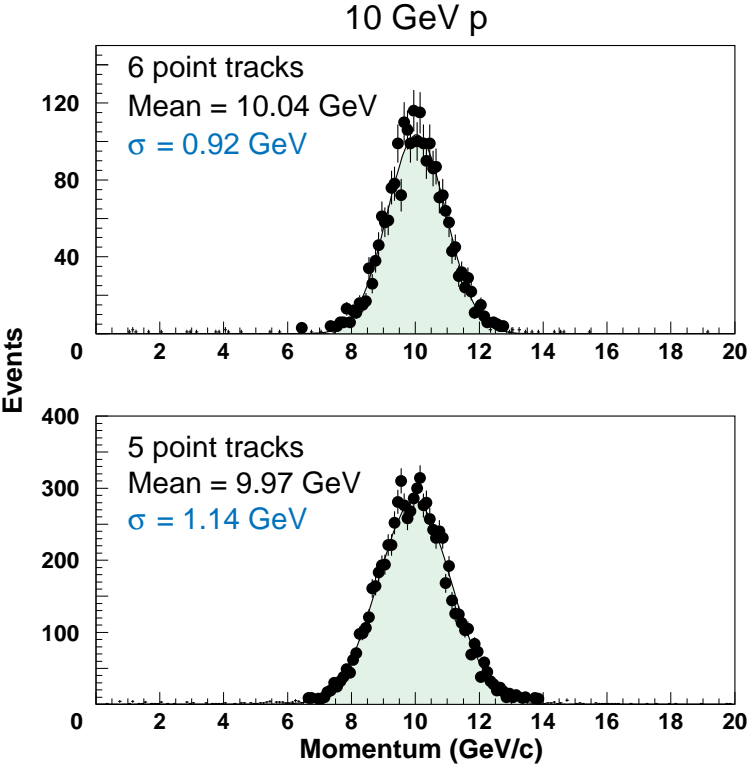,width=10 cm}}
\caption{Measurement of the Tracker momentum resolution at $10\ GeV$ using a proton  beam at CERN.
\label{fig:testbeam}}
\end{center}
\end{figure}

The curvature of the tracks is measured by up to six layers of silicon double
sided detectors, supported on ultralight honeycomb planes:  the total material 
traversed by a particle    is  very small,  $3.2\%$ of $X_0$ over the tracking volume and
 for normal incidence.
 The  momentum resolution
 of the  Silicon Spectrometer~\cite{ba}  is about $8\%$ 
in the region between $3$ and $10$ $GV$ of rigidity:
 at lower rigidities
 its  resolution worsen due to the multiple scattering
 while, at high energy,  the maximum detectable  rigidity ($\frac{\Delta R}{R}= 100\% $)
 is about $500\ GV$. The Tracker rigidity resolution function  was measured at the  GSI ion 
accelerator facility in Darmstadt in october 1998, using $He$ and $C$ beams, and at CERN in november
of the same year, using  a  proton  beam. The  results  confirm the design value; an example
  of the measured resolution is shown in Figure~\ref{fig:testbeam}. 
 The parameters of the  Silicon 
Spectrometer are given in Table~\ref{tab:sil}: 
 about $45\%$ of the   Tracker sensitive area was equipped  
during the precursor  flight, with a corresponding reduction on the spectrometer acceptance. 

\begin{figure}[htb]
\begin{center}
\mbox{\epsfig{file=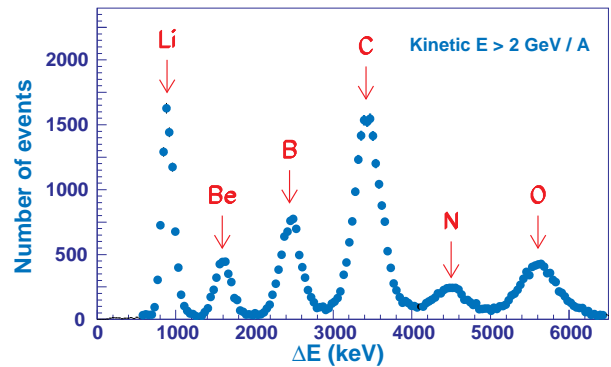,width=8.0cm}}
\caption{Energy deposition of the light ions as measured by the $ToF$ and Silicon Tracker systems.
\label{fig:ions}}
\end{center}
\end{figure}

Both the $ToF$ scintillators and the Silicon Tracker layers  measure   $\frac{dE}{dx}$, allowing a 
multiple determination of the absolute value of the particle charge, Z. 
  Figure~\ref{fig:ions} show the measurement of the energy deposited by 
different light nuclei during the precursor flight. 

\begin{table}[t]
  \caption{AMS silicon Tracker parameters (in parenthesis the values used on  the
precursor flight).  \label{tab:sil}}
 \begin{center}
  \begin{tabular}{|c|c|} \hline
   Number of planes &  8   \\
																		 & (6)  \\
 Accuracy (bending plane) & $10\  \mu m$  \\
Accuracy (non bending plane) & $30\  \mu m$  \\
Number of channels & 163936   \\
																		 & (58368)  \\
Power consumption & $400 \ W$ \\  
                  & ($180 \ W$) \\  
Weight & $130 \  kg$   \\
Silicon Area (double sided) & $5.4 \ m^2$   \\  
 																										 & ($2.4 \ m^2$)   \\ \hline
  \end{tabular}
 \end{center} 
\end{table} 
 \clearpage

\begin{figure}[htb]
\begin{center}
\mbox{\epsfig{file=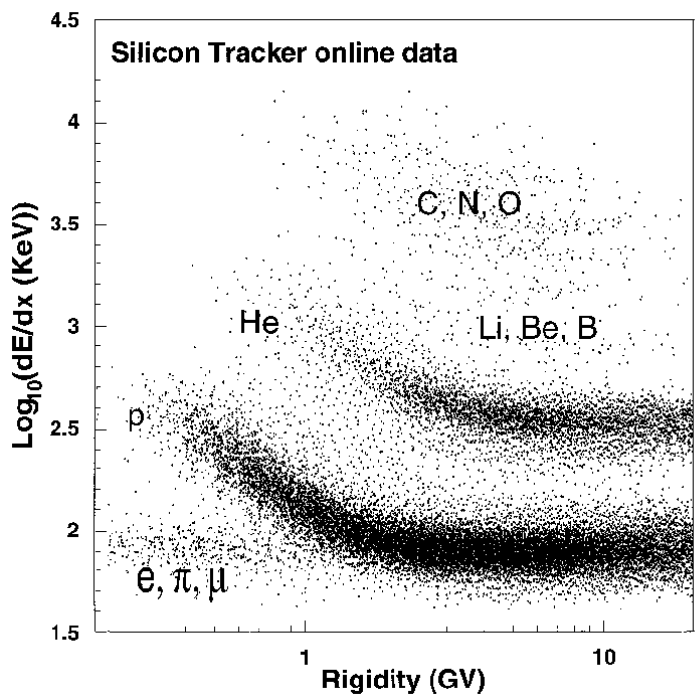,width=10cm}}
\caption{STS91 AMS flight online monitoring data. Both    dE/dx  and $|R|=|p/Z|$  
  are measured with  the Silicon Tracker.   \label{fig:De/Dx}}
\end{center}
\end{figure}

\begin{figure}[htb]
\begin{center}
\mbox{\epsfig{file=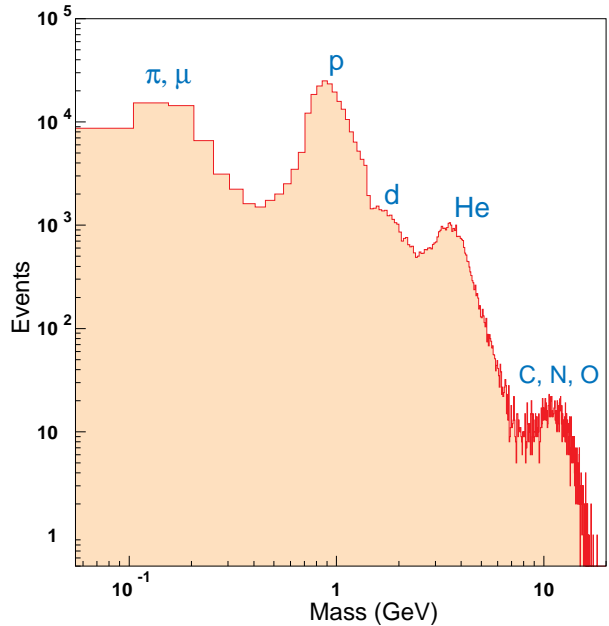,width=8cm}}
\caption{STS91 AMS flight online  monitoring data. Mass spectrum obtained by 
the combined measurement of the Silicon Tracker ($R$) and the Time of Flight ($\beta$).
 \label{fig:mass}}
\end{center}
\end{figure}

	All detector elements during production have undergone thermo-vacuum  tests  which 
have demonstrated  that neither the deep vacuum nor  important temperature variations deteriorate
detector performances. Systematic  vibration  tests  qualified 
that  the mechanical  design and workmanship were suited to stand 
the mechanical stresses during launch  and landing.

During the STS91 mission the  Spectrometer  collected data   at     trigger rates  varying from
 $\sim 100 ~Hz$ at the equator to $\sim 700~Hz$ at $\pm 52^o$, where
the event   rate was limited by the data acquisition speed. 

After  preprocessing and compression, the data were  stored
 on hard disks located on the Shuttle. A total of about 100 million triggers have been recorded
during the ten days mission.   A considerable part of the time, however, 
  the Shuttle  was docked to the  MIR station: 
in this condition the orientation was no good  for the AMS  since it was sometimes pointing 
 towards Earth. Besides, some element of the station were in the AMS view, thus producing
additional unwanted background. The useful time when only deep space was seen by the experiment
was about 4 days. 
Samples of the data ($< 10\%$ of the total) 
 were  are also sent to   ground in real time 
using  S-band  receiving ground  stations   at an average   rate of $1$ Mbit/s. 
Although  only  rough  calibrations  were  applied to these data, the reconstructed events
were   used   online  to  monitor  AMS operating conditions during the mission. 

For example, 	Figure~\ref{fig:De/Dx} shows the Tracker response to different types of CR during
the MIR docking  period: the double logarthmic plot  $dE\over{dx}$ versus $|R|=|{p\over{Z}}|$ clearly
shows bands corresponding  to light particles 
($e^\pm,\mu^\pm,\pi^\pm$), $p^\pm$, $\ ^3He/^4He$ and heavier  ions.
Figure  ~\ref{fig:mass} show  the   CR  mass spectrum obtained  from  $R$ measured by 
the Tracker   and $\beta$ measured by the ToF.

\begin{figure}[htb]
\begin{center}
\mbox{\epsfig{file=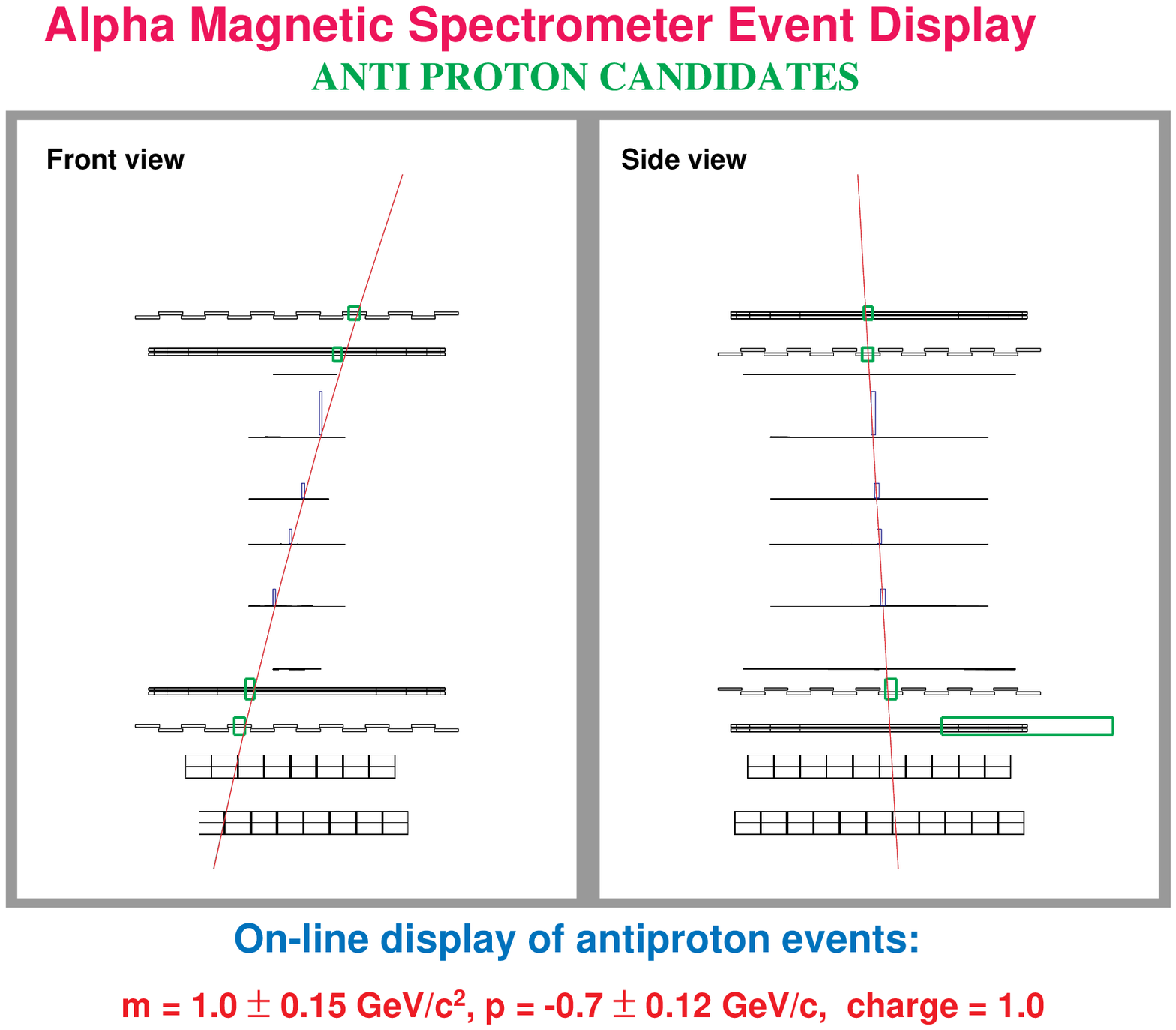,width=10cm}}
\caption{Candidate antiproton event from online monitoring data of the AMS precursor
flight. In the front view  the fraction of Silicon Tracker
equipped during the precursor flight is clearly visible. \label{fig:evento}}
\end{center}
\end{figure}

 A    candidate  $\bar{p}$ event,  measured online,  is also shown in  Figure~\ref{fig:evento}: 
one can note that the occupancy of the Silicon Tracker  is very  low, 
allowing unambiguous reconstruction 
of the particle trajectory in the magnetic field, meaning  the sign of its 
charge and  its momentum.   

	After landing, the full set of hard disks containing the data has been duplicated and the copy 
has been transported to CERN. During the month of august we   determined the in flight 
 calibration constants for the various detectors. The first mass production took place \
 in the fall of 1999  using a cluster of Alpha stations located at CERN.

\section{Antimatter Search}
 
 	To search for  nuclear antimatter, we search for particles with

\begin{itemize}
\item   negative rigidity;
\item   module of the charge  $Z$  equal or greater  than 2;
\item   mass   equal of greater than the He mass. 
\end{itemize}
These quantities are obtained through repeated measurements of the velocity  and its direction 
($ToF$ counters), the signed  momentum by the Silicon  Spectrometer,
 the absolute value of the charge from  $dE\over{dx}$ measurements of up to four  $ToF$ layers 
 and  up to 6 Silicon Tracker  layers.  

	We start with a preselection of  $Z>1$ particles and apply  soft quality cuts to reject  background 
particles with negative momentum  (${\bar{He}}$ and antinuclei with $Z>2$ candidates). 
The effects of these cuts 
is  studied on    control samples  containing  $5.7$ M  $He$ and   $276$ K $Z>2$ events.

\begin{itemize}
\item   
To    reject background due to single nuclear scattering in the Tracker we apply 
cuts on the particle rigidity $R$. $R$ is measured by the Silicon Spectrometer, 
using tracks having 5 or 6 points.  Since during the precursor 
 flight the Tracker  was only partially equipped, we inclued  in 
 this analysis  also events containing a track  detected only on     4 planes. 

	The particle rigidity is measured three times:  
the first two measurements $r^n_1$ and $r^n_2$  are obtained by using three consecutive  points 
out of the total number of measured points $n$, in the following way:  6 points patterns
  $r^6_{1}=r_{123}$, $r^6_{2}= r_{456}$, 5 points patterns  $r^5_{1}=r_{123}$, $r^5_{2}= r_{345}$,
 4 points patterns  $r^4_{1}=r_{123}$, $r^4_{2}= r_{234}$, where the lower indices  represent 
 the consecutive planes partecipating to the track fit.
  The third measurement,  $R$,  is obtained from a fit 
of all the points associated with one track.  In order to take properly   into account
the presence of multiple scattering,  we used  the  GEANE  fitting procedure~\cite{GEANT}.
 The three determination  
of the rigidity are compared  requiring that they give
 the same sign of the charge and consistent measurements of  the 
  momentum components. In particular, the comparison of the relative  rigidity error
 $\Delta R\over{R}$  with the  rigidity asymmetry $A_{12}=(r^n_1-r^n_2)/(r^n_1+r^n_2)$ 
allows the removal of  about $90\%$ of the negative momentum particles 
while keeping $79\%$ of the  $He$ control sample.

\item  To reject background due to an interaction of the primary particle
 in the Tracker material we apply  cuts on isolation  of the clusters detected on the
silicon planes. Events where too much  energy is observed within $5\ mm$ of the track 
are rejected. This cut reject  fifteen  times more particle in the sample with negative momentum: 
the positive momentum control samples are  basically unaffected 
($97\%$ of the events pass the cuts).

\item
To separate  between upward going and downward going particles we use the  $ToF$ measurement.

\item 
 The Identification of the absolute value of the particle charge is based on  
the repeated measurement of the $dE\over{dX}$  on the Silicon Tracker and $ToF$:
 we measure a contamination  between $p$ and $He$ below the  $10^{-7}$ level. 
\end{itemize}

\begin{figure}[htb]
\begin{center}
\mbox{\epsfig{file=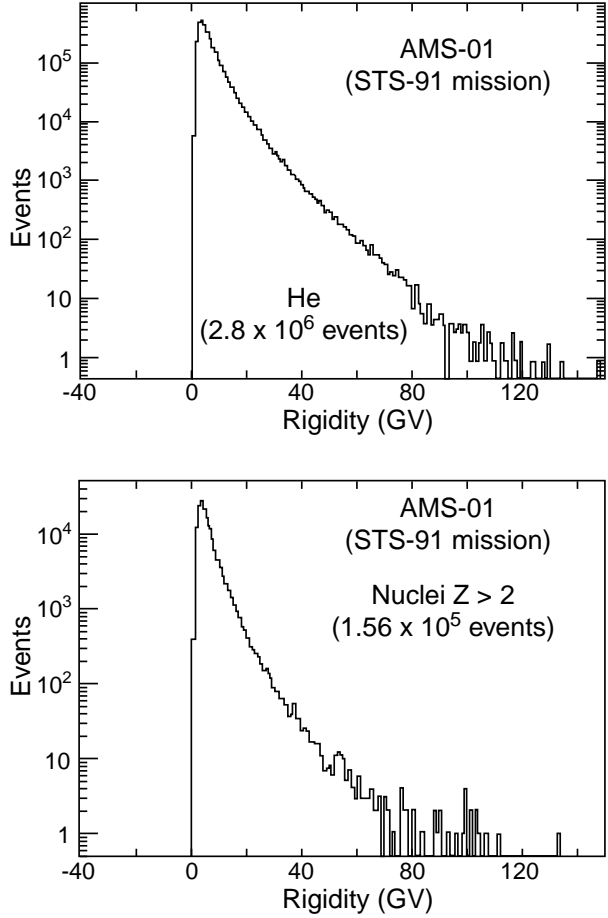,width=8.0cm}}
\caption{Light nuclei spectra  (control samples) as measured in the AMS precursor flight 
after having applied the antimatter selection cuts discussed in the text.
\label{fig:spectra}}
\end{center}
\end{figure}

\begin{figure}[htb]
\begin{center}
\mbox{\epsfig{file=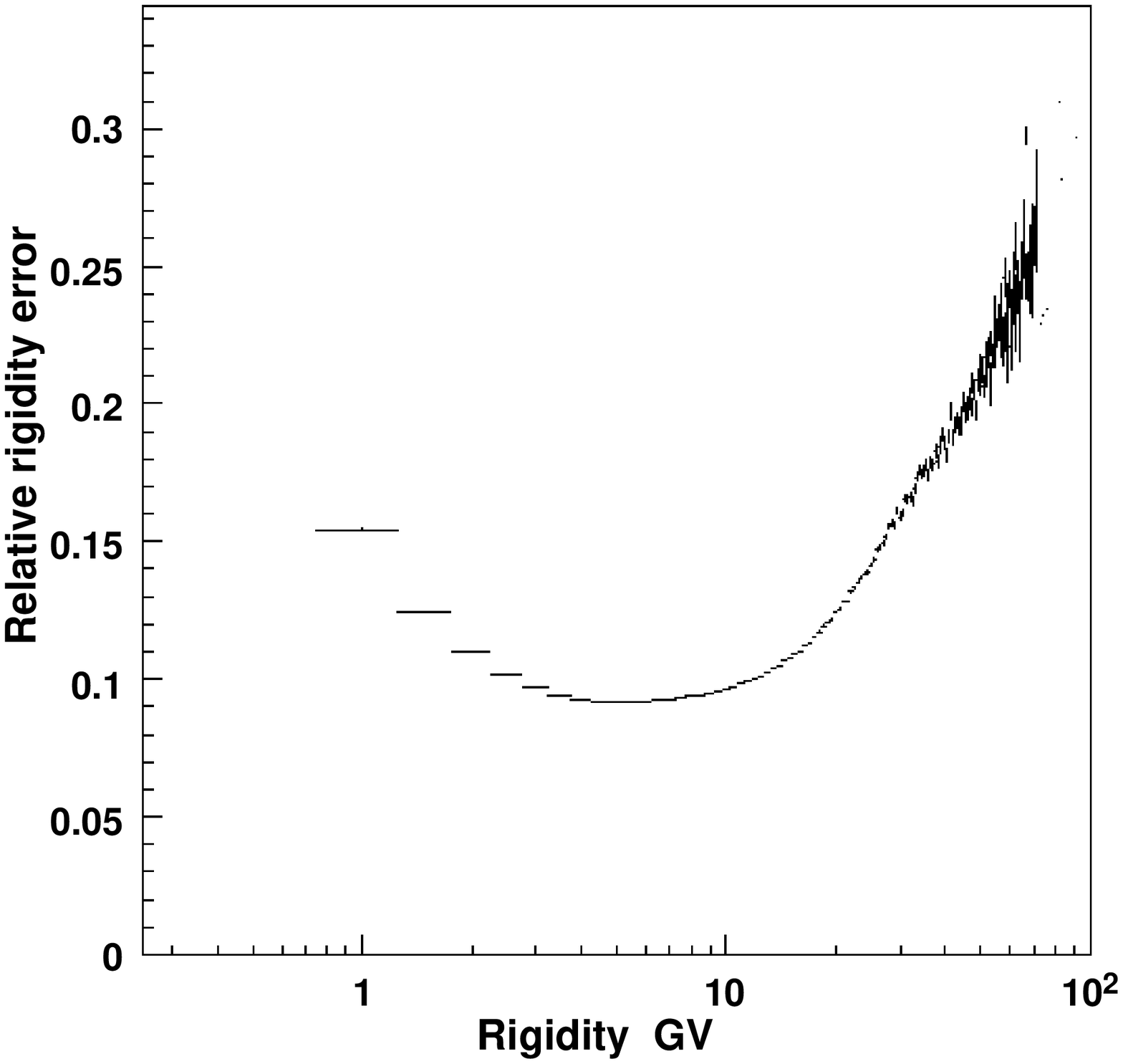,width=8.0cm}}
\caption{Momentum resolution of the  AMS  Silicon Tracker for $Z>1$ particles after having 
applied the antimatter selection cuts.
\label{fig:resolution}}
\end{center}
\end{figure}

	After the preselection we apply additional  $\chi^2$-cuts  on the track and $ToF$ measurements 
 and on an  overall likelihood  function describing the probability 
of an event to be compatible with   $He$, or heavier  nucleus, kinematics, mass and  velocity.
 Some of these cuts are  stricter for events hitting only 4 planes. 
	After these cuts all the candidates in the  $\bar{He}$ sample were removed while $2,8$ M
events on the $He$  sample survived, giving a total efficiency of about $49\%$.
 Similarly $156$ K events with $Z>2$  survived the
cuts, but none with negative  momentum, with a corresponding  cuts efficiency of about $56\%$.
The spectra of the positive charge samples after the cuts are shown in Figure~\ref{fig:spectra}: 
the spectra extends   above 100 GV of rigidity for both samples. 
The corresponding Tracker rigidity resolution is shown in
 Figure~\ref{fig:resolution}.

\section{Antimatter Limits}

	To establish a preliminary  antimatter upper limit    we proceed as follows. 
The flux of incident $He$ nuclei in a rigidity bin $(r, r+\Delta r)$  as a function of the measured
rigidity $r$, $\Phi_{He}(r)$, 
 is related to the measured He flux,  $\Phi^{M}_{He}(r)$, by

\be
	\Phi_{He}(r)=\epsilon^{-1}_{He}(r)\Phi^{M}_{He}(r)
\label{eq:s1}
\ee

where $\epsilon_{He}(r)$ is the rigidity dependent selection efficiency of
 the cuts discussed in the previous section,  simulated through  a complete MC simulation
using the  GEANT~\cite{GEANT} package. Trigger efficiency and the rigidity dependence 
of the anticounter  veto as well as the corrections due to electronics dead time
which was important on polar regions, was checked with events taken with an unbiased trigger. 
We also corrected  $\epsilon_{He}(r)$ for the $He-{\bar{He}}$ difference in absorption 
cross sections~\cite{moiseev}.  

	Since we detected no $\bar{He}$ candidate, the
differential upper limit  for the flux ratio at $95\%\ CL$ is given by:

\be
{{\Phi_{\bar{He}(r)}}\over{\Phi_{He}(r)}}< 
{{3 / \epsilon_{{\bar{He}}}(r)}\over{\epsilon^{-1}_{He}(r)\Phi^M_{He}(r)}} 
\label{eq:s2}
\ee

Since no $\bar{He}$ where found over all the measured rigidity range:

\be
\ \int\Phi^M_{\bar{He}}(r)dr <3  
\label{eq:s3}
\ee

\begin{figure}[htb]
\begin{center}
\mbox{\epsfig{file=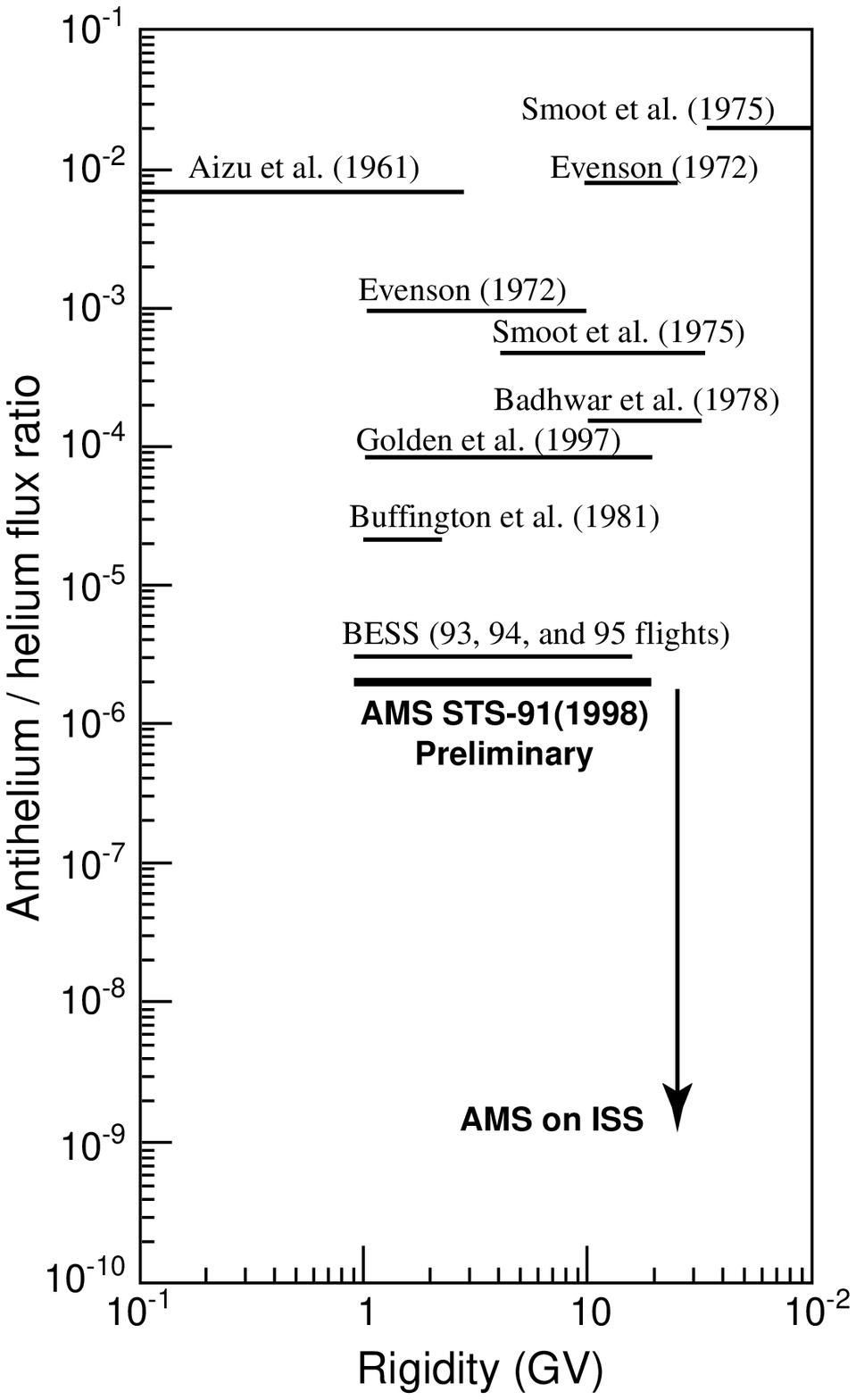,width=8.cm}}
\caption{Limits on $\Phi_{\bar{He}}\over{\Phi_{He}}$.
\label{fig:limits}}
\end{center}
\end{figure}

With the model dependent assumption that the $\bar{He}$ rigidity spectrum coincide with the 
$He$ spectrum we obtain:

\be
{{\Phi_{\bar{He}}}\over{\Phi_{He}}}< 1.14\ 10^{-6} 
\ee
Similarly with $Z>2$ data we obtain 

\be
{{\Phi_{\bar{Z>2}}}\over{\Phi_{Z>2}}}< 1.9\ 10^{-5}
\ee
We can also give a conservative  upper limit which does not depend on the unknown 
$\bar{He}$ energy spectrum~\cite{ag}. 
We integrate the arguments in equation (2) between   $r_{min}$  and $r_{max}$
 taking  the minimum value  of the efficiency in this rigidity interval
 $\epsilon^{min}_{{\bar{He}}}=min[\epsilon_{He}(r)]^{r_{max}}_{r_{min}}$. We calculate 

\be
{{\int^{r_{max}}_{r_{min}}\Phi_{\bar{He}}dr}\over{\int^{r_{max}}_{r_{min}}\Phi_{He}}dr}
< {{{3/ \epsilon^{min}_{{\bar{He}}}}}
\over{\int^{r_{max}}_{r_{min}}\epsilon^{-1}_{He}(r)\Phi^M_{He}(r)}dr}
\label{eq:s3}
\ee
which for  $r_{min}=1.6\ GV$  and $r_{max}= 20\ GV$  gives a model independent  limit on 
${{\Phi_{\bar{He}}}/{\Phi_{He}}}$ of $1.7\ 10^{-6}$ at $95\%$ of $CL$ while  
for $Z>2$  the corresponding  limit is  $ 2.8\ 10^{-5}$.
Figure~\ref{fig:limits} shows this  preliminary  result  for  ${\bar{He}}$
 compared with previous published results\cite{aa}$^{-}$\cite{ag} and the 
expected AMS sensitivity   on the ISS.
 Our result is better than the best limit  published   
by  BESS  adding the data of the '93, '94, and '95  flights at $56^o $  of latitude 
\cite{ag}. It also  spans over a larger rigidity interval. 
For  $Z>2$ our results is about 5 times better than the previous published 
results\cite{aa}$^,$\cite{ah}.
The large AMS acceptance  made possible  to set these stringent  limits using only
 4 days of exposition 
to deep  space.

\section {Conclusion}

 The AMS experiment has successfully completed the  first  precursor flight in June 1998 
with  an excellent performance of all subsystems,  collecting about 100 millions primary 
CR during 152 orbits around the earth. AMS   upper limits  on the existence  of antimatter 
improve the results of nearly 40 years of similar searches using stratospheric ballons. 
 
 There has never been a sensitive magnetic 
spectrometer in space  covering  the energy range  up to hundreds of GeV. 
	After its  installation on the ISS   in 2004, AMS will  measure   
the   CR rays composition with an accuracy orders of magnitude better than before. 
 This  instrument will open a  new sensitivity window in the search  for 
 antimatter and for  supersymmetric dark matter  in the galactic halo.

\end{document}